**Gas at Large Distances from Galaxies and QSO Absorption Lines**

Boqi Wang

*Department of Physics and Astronomy, The Johns Hopkins University, Baltimore, MD 21218*

# Gas at Large Distances from Galaxies and QSO Absorption Lines

Boqi Wang[1]

*Department of Physics and Astronomy, The Johns Hopkins University, Baltimore, MD 21218*

**Abstract.** We present the calculations of steady, spherical gas outflows from galaxies in an effort to understand the effects of the galaxy mass on the flow properties such as the size of the outflow regions, the efficiency of radiative cooling, and the fate of the cooled gas. We show that there exist no transonic flows but either subsonic or supersonic flows are obtainable, in an analogy with the stellar wind or accretion problems. Solutions of the supersonic outflows are studied in detail as they are most likely to carry gas to large distances away from galaxies. In case the gas does cool radiatively, the cooled gas is most likely to form clouds via various instabilities. The clouds coast farther away from the galaxy because of the finite kinetic energy they inherit. Depending on the initial energies, the clouds can either leave the galaxy or fall back ballistically. Applications of the calculations to dwarf and normal galaxies are made, and we argue that normal galaxies like our own should have relatively small gaseous halos formed through this mechanism. We discuss the implications of cold clouds from dwarf galaxies on recent observations of the QSO absorption line systems. For massive galaxies, satellites provide alternative ways to form large gaseous halos. We show how dynamical friction can increase tidal stripping of the gas as the satellite spirals in toward the primary, and that the gas ejected from satellites may account for the QSO heavy element absorption lines.

## 1. Introduction

Recent observations have increasingly shown evidence for the existence of gas associated with galaxies but at extremely large distances. Of great interest is the gas found through QSO absorption lines. For example, the QSO heavy element absorption line systems have long been suspected to reside in galactic halos (Bahcall & Spitzer 1969), and indeed recent attempts to identify spectroscopically the underlying galaxies that cause a subset of the absorption lines, namely the Mg II absorption lines at intermediate redshifts, have been successful (e.g., Bergeron & Boisse 1991; Steidel 1993). Those observations show that the gas causing Mg II absorption lines is associated with the galaxies at a typical distance of about 50 kpc from the galactic centers. Statistical studies of another subset of the absorption lines, the C IV absorption lines, indicate the regions of the absorbing gas are about twice as large as the Mg II systems (e.g., Sargent et al. 1988). More recently, studies of the low-column density Lyman $\alpha$ forests in QSO spectra in relation to the nearby galaxies suggest that they may be associated with galaxies, possibly at distances even larger than the heavy element absorption line systems (Weymann 1994; Lanzetta et al. 1994).

---

[1]Also Space Science Telescope Institute.

That the hot gas in the ISM heated by supernovae (SN) may flow out of the galaxy to large distances was first proposed by Shapiro and Field (1976), who examined the consequences of the hot and tenuous gas component in the interstellar medium (ISM), first put forward by Cox and Smith (1974). They noticed that the pressure of this hot gas would likely be larger than the then inferred pressure of the ISM, and without external confinement the gas would stream out the Galaxy. The gas could cool radiatively in the halo and subsequently fall back to the disk ballistically; hence the term "galactic fountain" was coined. Later McKee and Ostriker (1977) proposed that the ISM consists of three phases, with the most tenuous hot gas being dominant in volume. In their model, the hot gas will flow in to the Galactic halo, forming an X-ray halo. The galactic fountain model has subsequently been studied by means of hydrodynamic simulations by various authors (Bregman 1980, 1981; Habe & Ikeuchi 1980; Houck & Bregman 1990; Li & Ikeuchi 1992). Bregman (1981) proposed that the gas cooled in the halos of galaxies may account for the heavy element absorption lines seen in the QSO spectra, and Li and Ikeuchi (1992) calculated the ionization structures of the halo gas resulting from galactic fountains. Examinations of those hydrodynamic simulations reveal however that in galaxies like our own the heating rate required for the gas to reach a distance from the galactic center comparable to that observed in the metal line systems is implausibly large (Wang 1993). This is not surprising because the deep gravitational potential of galaxies like our own requires extremely energetic gas to move out of the potential. This may also explain the fact that ROSAT failed to detect such large X-ray halos around many normal type galaxies.

In an effort to understand how the hot gas can be transported from the galaxies to large distances, we here present the results of the cooling outflow solutions; a more detailed description can be found in Wang (1994). The present work is aimed at investigating in a systematic way the consequences of varying the mass of galaxies. To explore large parameter spaces, however, we make some simplifying assumptions. We assume that the outflow originates from a fixed initial (or base) radius at an initial temperature. We ignore thermal conduction, which can be shown to be unimportant in the large-scale dynamics of the upward flow since the heat flux conducted into the halo is usually smaller than the energy radiated (Houck & Bregman 1990). Any possible magnetic field (ignored here) will further impede thermal conduction. We assume spherically symmetric flows, which may be a reasonable approximation if the base of the flow is much smaller than the region of the outflow. We consider only steady outflows. The applications of the steady flows can be made to galaxies where the duration of the energy injection from SN into the ISM is sufficiently long, so the temperature at the base of the halo may be kept fixed during the flow, and a steady state can be achieved. Thus, the steady flow solutions can be applied to dwarf galaxies, where the outflow scales are relatively small, and disk galaxies like our own, where star formation has been relatively constant for the last few billion years (e.g., Wang & Silk 1994).

## 2. Formulation of the Problem

We consider a cooling rate per unit volume given by

$$C = A\rho^2 c^{2q} \tag{1}$$

where $c = (\gamma p/\rho)^{1/2}$ is the adiabatic sound speed, $\rho$ and $p$ are the gas density and pressure, and $\gamma$ is the ratio of the specific heat capacities. We assume that the circular (or rotation) velocity of a galaxy can be written as

$$v_{cir}^2 \equiv r\frac{d\phi}{dr} = \frac{B}{r^n} \tag{2}$$

where $\phi$ is the gravitational potential. For example, for a galaxy dominated by a dark halo with a flat rotation curve, $n = 0$ and $B = v_{cir}^2 = $ const. Given above, there exists a characteristic length, $r_0$, and a characteristic velocity, $v_0$ in steady spherical flows, where the mass loss rate $\lambda = 4\pi r^2 \rho v =$ const. (The gas density can be found from mass conservation; $\rho$ scales as $\rho_0 = \lambda/4\pi r_0^2 v_0$, and $p$ scales as $p_0 = v_0 \lambda/4\pi r_0^2$.) It is easy to show, from dimensional arguments using $A$, $B$, and $\lambda$, that they can be defined as the following:

$$r_0 = \left(\frac{A\lambda}{B^{2-q}}\right)^{1/(1-2n+nq)}$$

$$v_0 = \left(\frac{B}{A^n \lambda^n}\right)^{1/2(1-2n+nq)}. \tag{3}$$

For the logarithmic potential ($n = 0$), we obtain $r_0 = Av_{cir}^{2q}\lambda/v_{cir}^4$ and $v_0 = v_{cir}$. We can define dimensionless variables in the place of radius and sound speed (or temperature): $x = r/r_0$ and $w = c/v_0$. The dimensionless radius, $x$, is a parameter measuring the strength of the gravitational field of the galaxy relative to the radiative cooling rate. In fact, if $n = 0$ we have $x = \rho v_{cir}^2/4\pi(r/v_{cir})C$. That is, $1/x$ is the fraction of the energy lost to radiation during the flow time $r/v_{cir}$ for a flow with velocity equal to the rotation velocity. Therefore, if $x \gtrsim 1$ one expects that gravity becomes important in the flow. The dimensionless temperature, $w^2$ ($= c^2/v_{cir}^2$ for $n = 0$), is a parameter characterizing the gas thermal energy relative to the potential energy. Thus for $w^2 \gg 1$ we expect that the gas escapes the galaxy.

The equations of mass, momentum, and energy conservation can be combined to obtain the dimensionless equations for the Mach number and the dimensionless temperature:

$$\frac{dM^2}{dx} = \frac{(\gamma-1)x^n w^{2q}\left(1+\gamma M^2\right)+8\pi x^{1+n} w^4 M^2 \left[2+(\gamma-1)M^2\right]-4\pi(\gamma+1)xw^2 M^2}{4\pi x^{2+n} w^4 (M^2-1)}$$

$$\frac{dw^2}{dx} = \frac{(\gamma-1)\left[-x^n w^{2q}\left(\gamma M^2-1\right)-8\pi x^{1+n} w^4 M^4 + 4\pi x w^2 M^2\right]}{4\pi x^{2+n} w^2 M^2 (M^2-1)}. \tag{4}$$

The first, second, and third terms in the numerators of the equations above are due to contributions from radiative cooling, the gas pressure (or adiabatic expansion), and gravitation, respectively. The reduction to dimensionless form above greatly simplifies our investigation of the flow solutions in relation to various galaxy properties. The complexity of taking into account varying galaxy parameters are reduced to initial condition problems. Two crucial time scales in the present problem are the flow time and the cooling time. We define the flow time as $t_f = r/v$ and the cooling time as $t_c = p/(\gamma - 1)C$. The ratio of the cooling time to the flow time is then

$$\frac{t_c}{t_f} = \frac{4\pi}{\gamma(\gamma - 1)} x w^{4-2q} M^2. \tag{5}$$

The above ratio in fact is roughly the ratio of the contributions from adiabatic expansion and radiative cooling in supersonic flows in equation (4).

## 3. Supersonic Flows

In the absence of radiative cooling ($q \to -\infty$), and for the point-mass potential ($n = 1$), equation (4) reduces to the classic spherical accretion problem, first solved by Bondi (1952) for fixed adiabatic indices ranging from $\gamma = 1$ to $\gamma = 5/3$. The corresponding wind solutions have been discussed by Parker (1958) in the context of the hydrodynamic model of the solar wind. From these classic studies, it is well understood that solutions for transonic flows exist only for $\gamma < 5/3$. There are no solutions proceeding smoothly from subsonic to supersonic flow owing to the stiff pressure response of an adiabatic gas; the sonic point progressively moves toward the origin as $\gamma \to 5/3$. Solutions do exist, however, for either subsonic or supersonic flows for $\gamma \geq 5/3$. The branch of the solutions that are subsonic requires a finite confining pressure at infinity (Parker 1958).

For general potentials with arbitrary values of $n$, the topology of the solutions with constant $\gamma$ differs from the $n = 1$ case. We have obtained flow solutions for spherical accretion and wind problems at a fixed $\gamma$ (but without the cooling term included) in general potentials given by equation (2) (Wang 1994). We find that the range of $\gamma$ within which a transonic flow is possible is progressively reduced as $n$ decreases below unity. In particular, for $n = 0$, transonic flow can occur only if $\gamma < 1$. Above the critical value of $\gamma$ for given $n$ flows can only be either subsonic or supersonic. In subsonic outflows the density and temperature approach asymptotically constant values, implying a finite confining pressure at infinity, a generalization of the result found by Parker (1958) for the $n = 1$ case.

With radiative cooling ($q = $ finite) and $\gamma = 5/3$, pressure response to density variations is even steeper than in the adiabatic flow, and there exist no transonic flow solutions. Solutions do exist, however, for either supersonic or subsonic flows. In subsonic outflows, the gas are close to hydrostatic equilibrium, but the entropy of the gas decreases outward as a result of radiative cooling. Consequently, the subsonic outflow is subject to convective and thermal instabilities (Field 1965; Balbus & Soker 1989; Houck and Bregman 1990). The supersonic outflows, however, are driven by thermal pressure, and gravity is relatively unimportant for most of the flow. The gas becomes unstable only when it cools radiatively and gravity becomes important at the end of the flow (e.g., see Bregman 1980; Habe & Ikeuchi 1980; Li & Ikeuchi 1992). Furthermore, subsonic flows have higher densities than their supersonic counterparts. As a result, the radiative loss is more efficient, and the resulting halo size is smaller than in the supersonic flows. As we are mostly interested in the largest extent of the galactic halos resulting from outflows, below we shall concentrate on the supersonic flows.

The initial Mach number of the flow is taken as unity at the base of the flow, a natural choice for the initial condition in supersonic (or subsonic) outflows. Indeed, Chevalier and Clegg (1985) obtained an analytical solution for the wind driven from a region of uniform mass and energy deposition; both gravitational field and radiative cooling were ignored but a heating term was added to the flow equations. These conditions may be applicable to the region of the gas heating (the base). They found that a smooth transition from subsonic flow at the center to supersonic flow at large radius $r$ requires that $M = 1$ at $r = R$, where $R$ is the radius beyond which the heating ceases. The conditions imposed in the present work (i.e., inclusion of gravitational field and radiative cooling) are appropriate for the gas beyond the heating region, and thus our solutions can be considered as an extension of their inner ($r \leq R$) solutions; the transition is at $r = R$ with $M = 1$. We thus see that solutions of equation (4) are determined by two initial conditions $x_i$ and $w_i^2$ at the base radius, which reflect the relative importance of gravity to radiative cooling, and of the thermal energy to the potential energy of the initial gas.

## 4. General Properties of the Outflows

The characteristics of the outflows are determined by the ratio of the cooling time to the flow time, $t_c/t_f$, and by the initial dimensionless radius $x_i$. If the galactic gravitational potential is not important (i.e., $x_i \lesssim 1$), the outflows are distinguished by the initial values of $t_c/t_f$: (1) If initially $t_c/t_f \lesssim 1$, the gas cools as soon as it streams out of the galaxies. (2) If initially $t_c/t_f \gtrsim 1$, gas first cools adiabatically, and radiative cooling may become dominant later on. In case radiative cooling does become dominant, the gas temperature drops precipitously owing to the increase of the cooling rate with decreasing temperature. One may define a cooling radius $x_c$ where the gas temperature drops rapidly because of radiative cooling. Once radiative cooling sets in, the gas becomes unstable to convective and thermal instabilities (Balbus & Soker 1989) and the cloud formation ensues. The newly formed clouds drop out of the flow, and move ballistically. Because they inherit the kinetic energy of the gas at the point of cloud formation, they coast farther from the cooling radius. For large initial temperature and shallow potential wells, the clouds escape the galaxy, while for galaxies with deep potentials, the clouds remain in the halo.

The cooling radius $x_c$ for given initial dimensionless temperature $w_i^2$ is shown in figure 1 for several values of the initial dimensionless radius $x_i$, where we take $q = -0.6$ (appropriate for a gas of nearly cosmic abundance at $T \simeq 10^5 - 4 \times 10^7$ K) and $n = 0$ (a flat rotation curve). In the limit $w_i^2 \to 0$, the gas is trapped in the host galaxy, and $x_c$ is equal to $x_i$. As $w_i^2$ increases, $x_c$ increases, and asymptotically approaches $\propto w_i^{4.59}$ for large $w_i^2$. The transi-tion from gradual to the asymptotic rise of $x_c$ occurs at initial $t_c/t_f \sim 1$. The asymptotic behavior of $x_c$ can be understood as the following: Before the onset of radiative cooling the outflow is approximately adiabatic, so $M^2 \sim (x/x_i)^{4/3}$, and $w^2 \sim w_i^2 (x/x_i)^{-4/3}$. Radiative cooling becomes dominant when $t_c/t_f \lesssim 1$, which gives $x_c \propto x_i^{4(1-q)/(1-4q)} w_i^{6(2-q)/(1-4q)} = x_i^{1.88} w_i^{4.59}$. The cooled gas can coast farther beyond the cooling radius $x_c$ because the newly formed clouds inherit the finite kinetic energy from the hot gas. We show for given $w_i^2$ the maximum radius $x_f$ which the cold clouds can eventually reach (dotted curves), calculated for the same $x_i$. For definiteness, we have assumed that the mass distribution of the galaxies (and thus the flat rotation curve) is truncated at 50 times the initial radius, $x_{max} = 50 x_i$ (or $r_{max} = 50 r_i$).

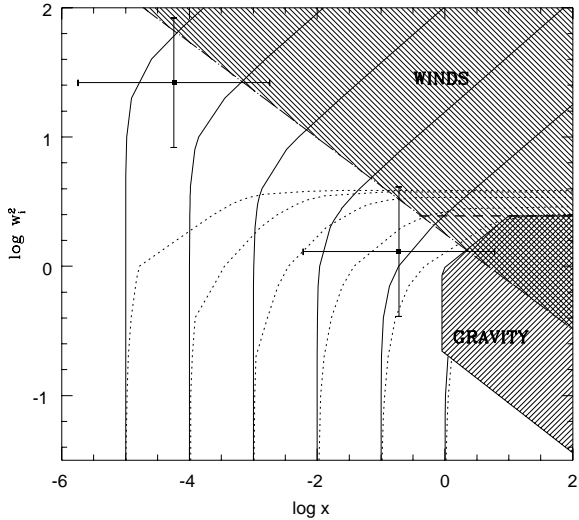

Figure 1. The initial dimensionless temperatures, $w = (c/v_0)^2$, versus the dimensionless radius, $x = r/r_0$. The cooling radius $x_c$ for given $w_i^2$ are calculated for the initial radius, from left to right, $x_i = 10^{-5}$, $10^{-3}$, $10^{-3}$, $10^{-2}$, $10^{-1}$, 1 (solid curves). The dotted curves denote the final radius $x_f$ which the cooled clouds can reach for the same $x_i$. The shaded area in the top right shows initial conditions $(w_i^2, x_i)$ where radiative cooling cannot be important. This is further divided into two regions; above the escape temperature (dashed horizontal line), the outflow results in galactic winds (denoted as WINDS), and below it the outflows lead to coronae. The other shaded area in the lower right denoted as GRAVITY represents $(x_i, w_i^2)$ where the gas is stopped by gravity before it has a chance to cool radiatively. The two crosses show the range of $(x_i, w_i^2)$ for dwarf galaxies with $v_{cir} = 50$ km/s (left cross) and normal galaxies with $v_{cir} = 225$ km/s (right cross).

If the temperature drop through adiabatic expansion is too large, e.g, below $10^5$K, before the radiative cooling can set in, the gas cannot cool radiatively because the radiative cooling time is increased at the low temperatures. This is equivalent to the requirement that initial $t_c/t_f$ be not too large. For example, for a gas with the initial temperature $T_i = 10^7$ K, it can be shown that the required initial ratio is $t_c/t_f \lesssim 42$. We plot $w_i^2$ above which radiative cooling cannot become dominant for a gas with $T_i = 10^7$K (shaded area above the dash-dotted line). If $T_i = \alpha \times 10^7$ K, the corresponding limit on $w_i^2$ changes by a factor of $\alpha^{(1-4q)/4(2-q)} = \alpha^{0.33}$. The region of $(x_i, w_i^2)$ that results in wind (denoted as WINDS) is bounded below by the requirement that $w_i^2$ be above the escape temperature. In the absence of radiative cooling the initial temperature necessary for the gas to escape is $w_i^2 = 2.456$ (dashed horizontal line) for $x_{max} = 50x_i$. Below the wind region the gas cannot leave the galaxy, resulting in a hot corona.

For $x_i \gtrsim 1$, gravity is strong relative to radiative cooling, so the gas will be pulled back to the galaxy before it has a chance to cool radiatively. We show the region of $(x_i, w_i^2)$ where the gas is confined to the halo by gravitation and radiative cooling is not important during the flow time (denoted as GRAVITY). The region is bounded below by $t_c/t_f \sim 1$ since obviously radiative cooling must not be too strong for gas to rise at all. It is also bounded above by the escape temperature. It is important to note that $t_c/t_f \sim 1$ at the base cannot be used as the sole criterion for the onset of the radiative cooling; adiabatic expansion brings the gas temperature down, so $t_c/t_f$ decreases in the flow. Only when $t_c/t_f \gg 1$ can radiative cooling be ignored.

## 5. Dwarf and Normal Galaxies

We now switch to dimensional variables, and make attempts to connect with observations. We take the cooling rate to be $A = 5.7 \times 10^{34}$ (in cgs) from Raymond et al. (1976), and $n = 0$. The initial gas temperature is primarily determined by the heating of the ISM by SN in the galaxy, which depends on the internal star formation rate per unit volume. Observationally, the star formation rate is found to correlate more with morphological types rather than the mass of galaxies (e.g., Sandage 1986), thus one may expect that $T_i$ is not a strong function of the galaxy mass. However, $x_i$ is a strong function of the circular velocity, thus the mass of the galaxy. As a result, for a given mass loss rate, massive galaxies in general have large $x_i$ and small $w_i$, and dwarf galaxies in general cluster around small $x_i$ and large $w_i$. This is expected if we recall that $x$ characterizes the strength of the gravitational field relative to the radiative cooling rate, and $w$ measures the thermal energy relative to the potential energy.

As examples, in figure 1 we show the inferred $(x_i, w_i^2)$ for dwarf galaxies with $v_{cir} = 50$ km/s (left cross). To take the uncertainties of the various parameters into account and to include dwarf galaxies in general, we allow for a generous range in the gas parameters: the vertical extension represents the adopted temperature range $T_i = 10^{6.5 \pm 0.5}$, and the horizontal stretch reflects the adopted range of ratio $r_{i,kpc}/\lambda_1 = 10^{\pm 1.5}$ (e.g., $\lambda_1 = 0.01 - 10$ for fixed $r_{i,kpc} = 0.3$) (left cross), where $r_{i,kpc}$ is the initial radius $r_i$ in kpc, and $\lambda_1 = \lambda/(M_\odot/\text{yr})$. (Note $x$ depends only on ratio $r/\lambda$.) Similarly, we also present the initial conditions expected for normal galaxies like our own with $v_{cir} = 225$ km/s (right cross); similar ranges for the gas parameters are assumed: $T_i = 10^{6.5 \pm 0.5}$ K, and $r_{i,kpc}/\lambda_1 = 10^{\pm 1.5}$ (e.g., $\lambda_1 = 0.1 - 100$ for fixed $r_{i,kpc} = 3$).

A comparison with the calculated cooling radius $x_c$ and the final radius $x_f$ for the cooled clouds in the figure shows that dwarf galaxies in general should have halos of hot gas many times their initial radii. Depending on the initial temperature and density, the gas can either cool or result in galactic winds. In either cases, the gas is most likely to leave galaxies because the temperature is usually above the escape temperature. In contrast, for massive galaxies like our own, unless the temperature is extremely high, the gas is in general confined within the galaxy. If the gas density is high enough (large $\lambda$ and thus small $x_i$), radiative cooling becomes important at the end and the gas cools in the halo. But typically the cooled gas is unable to travel much farther from the cooling radius because of the strong potential. If the gas temperature is below the escape temperature, and the density is low, the gas will be pulled back to the galaxy before it has a chance to cool radiatively. In this case, the flow will result in a galactic corona.

## 6. QSO Absorption Lines by Dwarfs

The above results may have significant implications for observations of the QSO absorption line systems. The huge size of the absorbing gas associated with the galaxies is rather hard to understand in the context of the outflows from normal galaxies like our own (Wang 1993). As seen in NGC891 and from many other failed attempts to detect the halo X-ray emissions, large halos are rare phenomena rather than the norm for normal galaxies. Indeed, estimates of the gas temperature of the hot phase in the ISM of the Galaxy show that the temperature is $T_i \lesssim 10^6$ K (McKee & Ostriker 1978). If we take $v_{cir} = 225$ km/s, $r_i = 3$ kpc (about one scale length) and $\lambda = 1$ M$_\odot$/yr, we have $(x_i, w_i^2)=(0.4,0.4)$ for galaxies like our Galaxy and NGC891. From figure 1 we see that the gas is trapped in the halo within a few kpc (also see Houck and Bregman 1990).

Less massive galaxies, in particular dwarf galaxies, may however provide the needed gas at large distances away from galaxies. The gas temperature of the hot phase in less massive galaxies may be comparable to normal galaxies because the same heating mechanism through the shocks of SN remnants is at work. But their gravitational potentials are much shallower than normal galaxies, therefore the gas is most likely to leave the galaxies. Our present work shows that the outflows from dwarf galaxies can cool radiatively for reasonable mass loss rates (e.g., a few tenths M$_\odot$/yr). The cooled gas is most likely to leave the galaxies as it inherits the kinetic energy of the hot gas which may not be exhausted by the cooling. The cooled gas thus can cause absorption over a area much larger than the original size of the galaxies. A more detailed discussion will be presented elsewhere.

## 7. Gas around Massive Galaxies

Two simple requirements for the outflows may further illustrate that gas outflows from normal, massive, galaxies like our own may not make up the large halos responsible for the QSO heavy element absorption lines (such as Mg II and C IV): (1) initial $t_c/t_f < f$ so the gas can cool radiatively, where $f$ is a numerical factor that can be found from flow solutions (cf. figure 1), and (2) $T_i$ has to be comparable to the escape temperature so the gas can reach large distances (50-100 kpc). Consider a outflow with initial radius $r_{i,kpc}$, temperature $T_i = T_6 \times 10^6$ K and sound speed $c_i$. Since a fluid element carries an amount of the internal energy per unit mass $E_i = (\gamma + 1)c_i^2/2(\gamma - 1) = 2c_i^2$, from equation (5) we find that requirement (1) implies an energy injection rate into the outflow

$$u = E_i \lambda > \frac{36\pi}{5} \frac{r_i c_i^{6-2q}}{fA} = 5 \times 10^{39} f^{-1} r_{i,kpc} T_6^{7.2} \quad \text{erg/s}. \tag{6}$$

Our calculations in §4 show that $f \simeq 20 T_6^{0.33}$. If we take the typical escape temperature to be $6 \times 10^6$ K (cf. in the solar neighborhood) for galaxies like our own, we find $u \gtrsim 5 \times 10^{43} r_{i,kpc}$ erg/s. This is equivalent to about 5 SN per year within 3 kpc of the galaxy *if the efficiency of the gas heating by SN is 100%*— an even higher SN rate if the efficiency is lower. In comparison, the inferred SN rate in galaxies like our Galaxy is about 0.02 per year (van den Bergh & Tammann 1991).

Can massive galaxies then possess large halos of gas at all? To answer this question, it is worth looking back at nearby galaxies. Our massive companion M31 is orbited by NGC205 and M32 each with about $10^9 M_\odot$ and a projected distance of 5-10 kpc. The Galaxy also has many satellites including LMC and SMC at a distance $\sim 50$ kpc. In fact, surveys of more distant galaxies find that satellites are quite common around normal spiral galaxies (Holmberg 1969; Zaritsky et al. 1993). Clearly, more massive galaxies have bigger gravitational influence on the surroundings, and therefore capture smaller galaxies as their satellites. The captured dwarf galaxies have their own internal star formation, and the resulting heated gas outflows from the galaxies, just as isolated dwarf galaxies discussed above. Once the gas streams out, it is under the influence of the primary's gravitational field, and stays in the halo of the primary.

More significantly, once the dwarf galaxies settle in the halo of the primary, they suffer from dynamical friction, a result of scatterings by mass particles in the halo off the satellites. Dynamical friction drags the satellite to spiral down to the center of the primary, and in the process gas is progressively stripped out of the satellite. For $n = 0$ potentials, both the rate of spiral-in of the satellite and the mass loss rate from the satellite only depend on the circular velocities of the primary and the satellite, $v_p$ and $v_s$ (Wang 1993):

$$\begin{aligned} \frac{dR}{dt} &\simeq -0.9 \frac{v_s^3}{v_p^2} \\ \frac{dm_s}{dt} &\simeq -0.6 \frac{v_s^6}{G v_p^3}, \end{aligned} \tag{7}$$

where $G$ is the gravitational constant, $R$ is the separation between the satellite and the primary, and $m_s$ is the mass of the satellite. For $v_s = 50$ km/s and $v_p = 225$ km/s, the rate of mass stripping is about 0.2 $M_\odot$/yr. So over about one orbital time, the mass accumulated in the primary's halo is about $2 \times 10^8 \, M_\odot$, roughly what is needed to account for the Mg II absorption line gas. Furthermore, since the decay rate of the satellite orbit is constant, the density of the gas stripped from the satellite is distributed roughly as $r^{-2}$ in the primary's halo. The rapid decrease of the gas density toward large radii may account for the difference in the inferred sizes of the Mg II and C IV absorption line systems; self-shieding of the UV ionizing radiation becomes unattainable once the density drops too low, and consequently the high ionization states of the heavy elments are more abundant in the outer part of the halo.

Thus, in this picture, the large mass of the galaxy will not hinder the formation of a large halo but is crucial to the creation of one. It also alleviates the problems of implausibly large star formation rate as in the outflow problems by using external satellites.

## References

Bahcall, J.N., &, Spitzer, L. 1969, ApJ, 156, L63.
Balbus, S.A., Soker, N. 1989, ApJ, 341, 611.
Bergeron, J., & Boisse, P. 1991, A&A, 243, 344.
Bondi, H. 1952, MNRAS, 112, 195.
Bregman, J.N. 1980, ApJ, 236, 577.
Bregman, J.N. 1981, ApJ, 250, 7.
Chevalier, R.A., & Clegg, A.W. 1985, Nature, 317, 44.
Cox, D.P. & Smith, B.W.1974, ApJ, 189, L105.
Field, G.B. 1965, ApJ, 142, 153.
Habe, A., & Ikeuchi, S. 1980, Prog. Theor. Phys, 64, 1955.
Holzer, T.E., & Axford, W.I. 1970, ARA&A, 8, 31.
Holmberg, E. 1969, Ark. Astr., 5, 305.
Houck, J.C., & Bregman, J.N 1990, ApJ, 352, 506.
Lanzetta, K.M., Bowen, D.V., Tytler, D., & Webb, J.K. 1994, submitted to ApJ.
Li, F., & Ikeuchi, S. 1992, ApJ, 390, 403.
McKee, C.F., & Ostriker, J.P. 1977, ApJ, 218, 148.
Parker, E.N. 1958, ApJ, 128, 664.
Raymond, J.C., Cox, D.P., & Smith, B.W. 1976, ApJ, 204, 290
Sandage, A. 1986, *Astr. Ap.*, **161**, 89.
Sargent, W.L.W., Boksenberg, A., & Steidel, C.C. 1988a, ApJS, 68, 5 39.
Shapiro, P.R., & Field, G.B. 1976, ApJ, 205, 762.
Steidel, C.C. 1993, in Evolution of Galaxies and Their Environment, Proceedings of the Third Teton Summer Astrophysics Conference, eds. J.M. Shull and H. Thronson (Dordrecht: Kluwer), in press.
van den Bergh, S. & Tammann, G.A. 1991, Ann. As. Ap, 29, 363.
Wang, B. 1993, ApJ, 415, 174.
Wang, B. 1994, to appear in ApJ.
Wang, B., & Silk, J. 1994, ApJ, 427, 759.
Weymann, R. 1994, this conference.
Zaritsky, D., Smith, R., Frenk, C., & White, S.D.M. 1993, ApJ, 405, 464.